\documentclass[conference]{IEEEtran}
\IEEEoverridecommandlockouts
\usepackage{cite}
\usepackage{hyperref}
\usepackage{amsmath,amssymb,amsfonts}
\usepackage{amsmath}
\DeclareMathOperator*{\argmax}{argmax}
\usepackage{graphicx}
\usepackage{textcomp}
\usepackage{xcolor}
\usepackage{graphicx}
\usepackage{float} 
\usepackage{subfigure}
\usepackage{amsmath}
\usepackage{amsfonts,amssymb}
\usepackage{mathrsfs}
\usepackage{mathtools}
\usepackage{algorithm}
\usepackage{algorithmicx}
\usepackage{algpseudocode}
\usepackage{stfloats}
\bibliography{IEEEabrv}
\def\BibTeX{{\rm B\kern-.05em{\sc i\kern-.025em b}\kern-.08em
    T\kern-.1667em\lower.7ex\hbox{E}\kern-.125emX}}

\begin{document}

\title{Asymptotic Upper Capacity Bound for Receive Antenna Selection in Massive MIMO Systems\\
}

\author{\IEEEauthorblockN{Chongjun Ouyang, Zeliang Ou, Lu Zhang and Hongwen Yang}
\IEEEauthorblockA{\textit{Wireless Theories and Technologies Lab} \\
\textit{Beijing University of Posts and Telecommunications}\\
Beijing, China \\
\{DragonAim, ouzeliang, zhanglu\_96, yanghong\}@bupt.edu.cn}
}

\maketitle

\begin{abstract}
This paper studies the receive antenna selection in massive multiple-input multiple-output (MIMO) system. The receiver, equipped with a large-scale antenna array whose size is much larger than that of the transmitter, selects a subset of antennas to receive messages. A low-complexity asymptotic approximated upper capacity bound is derived in the limit of massive MIMO systems over independent and identical distributed flat fading Rayleigh channel, assuming that the channel side information (CSI) is only available at the receiver. Furthermore, the asymptotic theory is separately applied to two scenarios which is based on whether the total amount of the selected antennas exceed that of the transmit antennas. Besides analytical derivations, simulation results are provided to demonstrate the approximation precision of the asymptotic results and the tightness of the capacity bound.
\end{abstract}


\section{Introduction}
\label{sction1}
Massive multiple-input multiple-output (MIMO) system drastically improves the spectral efficiency by deploying large-scale antenna arrays at the base station (BS) \cite{b13,b21} and thus is deemed to be one of the most prospective approaches in the 5th generation cellular networks (5G) and {\rm{THz}} communication. In addition, the work in \cite{b14} showed the significant improvements of transmission security and reliability in massive MIMO channels compared with the small-scale system. To promise communication, each antenna should be connected with a radio-frequency (RF) chain, which results in high hardware cost in Massive MIMO system. In this respect, massive MIMO system with antenna selection  (AS-MIMO) {\cite{b2}} has gained significant attentions in recent years aiming for design of high-efficiency transmission schemes \cite{b15,b20,b21}.

Antenna selection (AS) technology \cite{b2} is regarded as an alternative to alleviate the requirement on the RF transceivers by selecting a subset of antennas to transceive signals. At the BS, antenna  selection has been applied into massive MIMO channels for both uplink and downlink transmission. \cite{b33} firstly defined the upper capacity bound of AS-MIMO to measure its performance analytically.  
\cite{b30} analyzed the channel capacity of AS-MIMO system in the limit of large and small Signal to Noise Ratio (SNR). The work in \cite{b22,b34} analyzed the performance of antenna selection under imperfect channel side information (CSI). In addition to performance analysis, many algorithms for antenna selection have been proposed based on different performance criterion, such as channel capacity \cite{b12,b17,b19,b3,b23} and bit error rate (BER)\cite{b25}.    

Asymptotic theory on order statistics \cite{b28,b32} can be applied into massive MIMO systems to simplify some derivations or approximate some system performance due to the properties of large dimensionality originated from large-scale antenna arrays. By the asymptotic theory, \cite{b31} derived the approximate distribution of channel capacity of MIMO systems over Rayleigh channels. Upper capacity bound in AS-MIMO was first proposed in \cite{b33}, and the exact expressions for it was derived. \cite{b3} simplifies the derivations in \cite{b33} slightly using asymptotic theory in massive MIMO channels. The work in \cite{b23} utilized asymptotic theory to analyze the norm-based antenna selection algorithm and obtained an excellent approximation to the channel capacity. Furthermore, \cite{b26,b27} extended \cite{b23} to massive Multiple-Input Multiple-Output Multiple-Eavesdropper (MIMOME) channels \cite{b29} to explore the relationship between the number of RF chains and transmission security . 

This paper concentrates on channel capacity for receive antenna selection (RAS) in massive MIMO systems. The asymptotic form of the upper bound is derived based on asymptotic theory with the guarantee of approximation precision. Intuitively, the computation complexity of this approximation result is much lower compared with those in \cite{b33} and \cite{b3}. For simplicity, suppose that the CSI is unavailable at the transmitter and the total transmit power is uniformly allocated. By the definition of upper capacity bound \cite{b33}, the asymptotic approximation is discussed in two scenarios : 1) For Scenario A: the number of the selected antennas is no larger than that of the transmit antennas, and 2) For Scenario B: the amount of the selected antennas exceed that of the transmit antennas. In each scenario, simulation results demonstrate that the derived asymptotic bound has good approximation effect. 

The remaining parts of this manuscript is structured as follows: Section \ref{sec2} describes the system model. In Section \ref{sec3}, the asymptotic upper bound is derived. The simulation results and corresponding analysis are shown in Section \ref{sec5}. Finally, Section \ref{sec6} concludes the paper.      

$Notations$: Scalars, vectors and matrices are denoted by non-bold, bold lower case, and bold upper letters, respectively. $\mathbb{C}$ stands for the complex numbers. The Hermitian and inverse of matrix $\bf{H}$ is indicated with ${\bf{H}}^{\dagger}$ and ${\bf{H}}^{-1}$, and ${\bf{I}}_N$ is the $N{\times}N$ identity matrix.

\section{System Model} 
\label{sec2}       
In this paper, we consider a massive MIMO system in which the transmitter ie equipped with $N_{\rm{t}}$ antennas and the receiver is equipped with $N_{\rm{r}}$ antennas. The received signal vector at the receiver reads
\begin{equation}
{{\bf{y}}}=\sqrt{{\rho}}{{\bf{H}}}\bf{x}+{{\bf{w}}},
\end{equation}
where ${\bf{x}}\in{\mathbb{C}}^{N_{\rm{t}}\times1}$ is the transmitted signal with unit power, ${{\rho}}$ is the SNR at each receive antenna, and ${{\bf{w}}}{\in}{\mathcal{CN}}{({\bf{0}},{{\bf{I}}_{N_{\rm{t}}}})}$ is the additive complex Gaussian noise. Assume that the transmitted symbols from different antennas are independent. Considering independent and identically distributed (i.i.d) Rayleigh flat fading channel, the elements in channel matrix ${\bf{H}}{\in}{{\mathbb{C}}^{{N_{\rm{r}}}{\times}{N_{\rm{t}}}}}$ are i.i.d. complex Gaussian random variables following ${\mathcal{CN}}(0,1)$. Suppose that the channel side information is only available at the receiver and the transmit power is uniformly allocated, the channel capacity can be written as \cite{b1}
\begin{equation}
{C}=\log_2\det\left({{\bf{I}}_{N_{\rm{r}}}}+\frac{{\rho}}{N_{\rm{t}}}{\bf{H}}{{\bf{H}}^{\dagger}}\right).
\end{equation}  

Then consider the RAS at the receiver and $L$ antennas are selected. Actually, selecting a subset of receive antennas is, in other words, to select the corresponding rows of channel matrix. Let ${{\tilde{\bf{H}}}}{\in}{{\mathbb{C}}^{{L}{\times}{N_{\rm{t}}}}}$ denote the submatrix after RAS, the corresponding channel capacity is represented as 
\begin{equation}
{\tilde{C}}=\log_2\det\left({{\bf{I}}_{L}}+{\overline{\rho}}{\tilde{\bf{H}}}{{\tilde{\bf{H}}}}^{\dagger}\right),
\end{equation}
where ${\overline{\rho}}=\frac{{\rho}}{L}$ is defined as the normalized SNR. Let $S$ denote the selected subset of receive antenna indexes whose cardinality is $\left|S\right|=L$, the goal of RAS is summarized as 
\begin{equation}
S^{\rm{opt}}=\argmax_{S{\in}{\mathcal{M}}}\log_2\det\left({{\bf{I}}_{L}}+{\overline{\rho}}{\tilde{\bf{H}}}{{\tilde{\bf{H}}}}^{\dagger}\right),
\end{equation} 
where $\mathcal{M}$ denotes the full set of all the candidate row index subsets with size $L$. Denote ${\tilde{\bf{H}}}_{\rm{s}}$ as the corresponding submatrix of $S^{\rm{opt}}$, the channel capacity reads
\begin{equation}
{C_{\rm{s}}}=\log_2\det\left({{\bf{I}}_{L}}+{\overline{\rho}}{\tilde{\bf{H}}}_{\rm{s}}{\tilde{\bf{H}}}{_{\rm{s}}^{\dagger}}\right).
\label{EQU5}
\end{equation} 

\section{Upper Bound}
\label{sec3}

It was virtually impossible to know the analytical solution to the optimal channel capacity $C_{\rm{s}}$ after RAS \cite{b33} for its prohibitive computation complexity stemming from exhaustive search (ES) especially when $L$ is large. Thus it makes sense to define the capacity upper bound to measure the performance of antenna selection technology. There are two types of capacity upper bound defined for antenna selection in MIMO system \cite{b33}. The first type is used when $L{\leq}N_{\rm{t}}$, in which the system is treated as $N_{\rm{r}}$ independent MISO subsystems. In each subsystem, beamforming (BF) strategy is used and the best $L$ ones of these subsystems are selected. The second one is used when $L{>}N_{\rm{t}}$, in which the system is treated as $N_{\rm{t}}$ independent SIMO subsystems and the best $L$ receive antennas for maximal ratio combination (MRC) in each subsystem are activated. We use BF Upper Bound (BUB) and MRC Upper Bound (MUB) to term these two bounds respectively.

\subsection{BF Upper Bound}
\label{sec4.1}
The interpretation of BUB originates from the definition of the upper capacity bound for the full-complexity MIMO system used in \cite{b1}, which reads
\begin{equation}
\label{equ5}
{C_{\rm{full}}}=\sum_{i=1}^{N_{\rm{t}}}\log_2\left(1+{\overline{\rho}}\alpha_i\right),
\end{equation} 
where $\{\alpha_i\}_{i=1,2,\cdots,N_{\rm{t}}}$ are independent chi-squared-distributed random variables with $2N_{\rm{r}}$ degrees of freedom. Equ.(\ref{equ5}) displays an artificial case when each of the $N_{\rm{t}}$ transmitted signals is received by a separate set of $N_{\rm{r}}$ receive antennas without interference from each other \cite{b1}. According to this explanation, a new upper capacity bound is defined by exchanging the roles of transmitter and receiver \cite{b33,b3}, which reads
\begin{equation}
\label{equ6}
{{\tilde{C}}_{\rm{full}}}=\sum_{i=1}^{N_{\rm{r}}}\log_2\left(1+{\overline{\rho}}\gamma_i\right),
\end{equation}   
where $\{\gamma_i\}_{i=1,2,\cdots,N_{\rm{r}}}$ are independent chi-squared-distributed random variables with $2N_{\rm{t}}$ degrees of freedom. The new definition still indicates an unrealistic situation when each of the $N_{\rm{r}}$ receive antennas has its own set of transmit antennas. By Equ.(\ref{equ6}), the upper capacity bound with antenna selection reads\cite{b33,b3}
\begin{equation}
\label{equ7}
{{\tilde{C}}_{\rm{s}}}=\sum_{i=1}^{L}\log_2\left(1+{\overline{\rho}}\gamma_{(i)}\right),
\end{equation}
where $\{\gamma_{(i)}\}_{i=1,2,\cdots,N_{\rm{r}}}$ are $ordered$ chi-square-distributed variables with $2N_{\rm{t}}$ degrees of freedom, i.e. $\gamma_{(1)}{\geq}\gamma_{(2)}{\geq}{\cdots}{\geq}\gamma_{(N_{\rm{r}})}$. It is artificial but can serve as an upper bound. The work in \cite{b33} proved that this bound is relatively tight when $L{\leq}N_{\rm{t}}$ holds. Nevertheless, the acquisition of the analytical form of the ${\tilde{C}}_{\rm{s}}$ is computationally complex, especially in the large-scale scenario when $N_{\rm{r}}$ is colossal\cite{b33,b3}. 

In sense of large-scale behavior, the asymptotic theory has become a topic of interest to alleviate computation complexity. Instead of calculating the exact joint distribution of the top-$L$ variables from  $\{\gamma_{(i)}\}_{i=1,2,\cdots,N_{\rm{r}}}$, the asymptotic theory derives an approximate distribution of them with properly high precision. In contrast to the analytical solution of $\tilde{C}_s$, more simplified computational expressions are available by asymptotic theory.

Actually, $\sum\limits_{i=1}^{L}\log_2\left(1+{\overline{\rho}}\gamma_{(i)}\right)$ is the sum of the top-$L$ $ordered$ statistics from $\left\{\log_2\left(1+{\overline{\rho}}\gamma_{(i)}\right)\right\}_{i=1,2,\cdots,N_{\rm{r}}}$, which is termed as a trimmed sum\cite{b4}. The distribution of a trimmed sum is shown to converge to be normal as the total size $N_{\rm{r}}$ tending to infinite\cite{b4}. Furthermore, the distribution of $\sum\limits_{i=1}^{L}\log_2\left(1+{\overline{\rho}}\gamma_{(i)}\right)$ converges rapidly with increment of $N_{\rm{r}}$, which is verified by simulation results in Section \ref{sec5}. Therefore, a normal approximation can be applied to the trimmed sum even though the range size $N_{\rm{r}}$ is of limited length. By the theorem in \cite{b4}, $\sum\limits_{i=1}^{L}\log_2\left(1+{\overline{\rho}}\gamma_{(i)}\right)$ is approximated as a Gaussian random variable $g{\sim}{{\mathcal{N}}\left(\mu_g,\sigma_g^2\right)}$. As $N_{\rm{r}}$ rises, the approximate error will converge to zero with fixed $L$. Based on the main theorem in \cite{b4}, $\mu_g$ and $\sigma_g^2$ are determined as
\begin{subequations}
\begin{align}
{\mu_g}&=N_{\rm{r}}\int_u^{\infty}\log_2\left(1+{\overline{\rho}}x\right){f_{N_{\rm{t}}}\left(x\right)}{\rm{d}}x\\
{\sigma_g^2}&=L\left(\sigma^2+\left(u-\frac{\mu_g}{L}\right)^2\left(1-\frac{L}{N_{\rm{r}}}\right)\right),
\end{align}
\label{equ8}
\end{subequations}
where 
\begin{equation}
{\sigma^2}=\frac{N_{\rm{r}}}{L}\int_u^{\infty}\left(\log_2\left(1+{\overline{\rho}}x\right)\right)^2{f_{N_{\rm{t}}}\left(x\right)}{\rm{d}}x,
\label{sigma}
\end{equation}
and ${f_{N_{\rm{t}}}\left(\cdot\right)}$ denotes the chi-squared probability density function (PDF) with $2N_{\rm{t}}$ degrees of freedom and mean $N_{\rm{t}}$ which reads\cite{b8}
\begin{equation}
{f_{N_{\rm{t}}}\left(x\right)}=\frac{1}{\left({N_{\rm{t}}}-1\right)!}\left\{\begin{aligned}&e^{-x}x^{N_{\rm{t}}-1},&x{\geq}0\\&0,&x{<}0\end{aligned}\right..
\label{equ9}
\end{equation}   
The constant $u$ in Equ.(\ref{equ8}) satisfies $\int_u^{\infty}{f_{N_{\rm{t}}}\left(x\right)}{\rm{d}}x=\frac{L}{N_{\rm{r}}}$ which can be solved 
by table-referring. Substitute Equ.(\ref{equ9}) into Equ.(\ref{equ8}) and Equ.(\ref{sigma}), $\mu_g$ and $\sigma^2$ are simplified after some derivations, which are exhibited on the top of the next page.
\newcounter{mytempeqncnt}
\begin{figure*}[!t]
\normalsize
\setcounter{mytempeqncnt}{\value{equation}}
\setcounter{equation}{11}
\begin{subequations}
\begin{align}
{\mu_g}&=N_{\rm{r}}\int_u^{\infty}\log_2\left(1+{\overline{\rho}}x\right){f_{N_{\rm{t}}}\left(x\right)}{\rm{d}}x=N_{\rm{r}}\int_u^{\infty}\log_2\left(1+{\overline{\rho}}x\right){\rm{d}}\left(-\sum_{k=0}^{N_{\rm{t}}-1}\frac{x^k}{e^xk!}\right)\notag\\&=\frac{{N_{\rm{r}}}}{\ln2}\sum_{k=0}^{N_{\rm{t}}-1}\ln\left(1+{\overline{\rho}}u\right)\frac{u^k}{e^uk!}+\frac{\overline{\rho}}{\ln2}\sum_{k=0}^{N_{\rm{t}}-1}\int_{u}^{\infty}\frac{{N_{\rm{r}}}x^k}{e^x\left(1+{\overline{\rho}}x\right)k!}{\rm{d}}x\\
{\sigma^2}&=N_{\rm{r}}\int_u^{\infty}\left(\log_2\left(1+{\overline{\rho}}x\right)\right)^2{f_{N_{\rm{t}}}\left(x\right)}{\rm{d}}x=N_{\rm{r}}\int_u^{\infty}\left(\log_2\left(1+{\overline{\rho}}x\right)\right)^2{\rm{d}}\left(-\sum_{k=0}^{N_{\rm{t}}-1}\frac{x^k}{e^xk!}\right)\notag\\&=\sum_{k=0}^{N_{\rm{t}}-1}{N_{\rm{r}}}\left(\log_2\left(1+{\overline{\rho}}u\right)\right)^{2}\frac{u^k}{e^uk!}+\frac{2\overline{\rho}}{\ln2}\sum_{k=0}^{N_{\rm{t}}-1}\int_{u}^{\infty}\frac{{N_{\rm{r}}}x^k\log_2\left(1+{\overline{\rho}}x\right)}{e^x\left(1+{\overline{\rho}}x\right)k!}{\rm{d}}x.
\end{align}
\label{equ10}
\end{subequations}
\setcounter{equation}{\value{mytempeqncnt}}
\hrulefill
\vspace*{4pt}
\end{figure*}

The integrals in Equ.(\ref{equ10}) can be solved efficiently using numerical integration owing to the attenuation of the term $e^{-x}\left(1+{\overline{\rho}}x\right)^{-1}$. The asymptotic approximation for the PDF of the trimmed sum $\sum\limits_{i=1}^{L}\log_2\left(1+{\overline{\rho}}\gamma_{(i)}\right)$ is written as
\setcounter{equation}{12}
\begin{equation}
p_{\rm{B}}\left(x\right)=\frac{1}{\sqrt{2\pi\sigma_g^2}}e^{-\frac{\left(x-\mu_g\right)^2}{2\sigma_g^2}}.
\label{pB}
\end{equation}

It is shown in \cite{b33} that the BF Upper Bound is relatively tight when $L{\leq}N_{\rm{t}}$. In addition, the less the number of selected antennas is, the tighter the upper bound is \cite{b33}. Considering the extreme case when $L=1$ in a manner where the MIMO system after RAS degrades into a MISO system, the BUB simply equals to the channel capacity $C_{\rm{s}}$ in Equ.(\ref{EQU5}). However, the upper capacity bound for the AS-MIMO should be rewritten when the number of activated antennas is larger than  $N_{\rm{t}}$.   

\subsection{MRC Upper Bound} 
\label{sec4.2}
The upper capacity bound when ${L}{>}{N_{\rm{t}}}$ is defined as \cite{b33}
\begin{equation}
{\tilde{C}}_{\rm{s}}=\sum_{h=1}^{N_{\rm{t}}}\log_2\left(1+{\overline{\rho}}\sum_{i=1}^{L}{\tilde{\gamma}_{(i)}}\right)=\sum_{h=1}^{N_{\rm{t}}}\xi_h,
\label{equ11}
\end{equation}
where $\{\tilde{\gamma}_{(i)}\}_{i=1,2,\cdots,N_{\rm{r}}}$ are $ordered$ chi-square-distributed variables with 2 degrees of freedom, i.e. $\tilde{\gamma}_{(1)}{\geq}\tilde{\gamma}_{(2)}{\geq}{\cdots}{\geq}\tilde{\gamma}_{(N_{\rm{r}})}$. Equ.(\ref{equ11}) presents a case when each of the $N_{\rm{t}}$ antennas communicates with a separate receive antenna subsets with size $N_{\rm{r}}$ in a manner where no interferences among these independent SIMO subsystems occur\cite{b33}. The best $L$ receive antennas are selected for maximal ratio combination in each subsystem, which also refers to hybrid
selection/maximum ratio combining (H-S/MRC) \cite{b5,b6}. Thus, the MRC Upper Bound for $L>N_{\rm{t}}$ in Equ.(\ref{equ11}) holds.

Following the similar steps in section \ref{sec4.1}, the trimmed sum $\sum_{i=1}^{L}{\tilde{\gamma}_{(i)}}$ can be asymptotically approximated as a Gaussian random variable $t\sim\mathcal{N}\left(\mu_t,\sigma_t^2\right)$ with mean and variance given as
\begin{subequations}
\begin{align}
{\mu_t}&=N_{\rm{r}}\int_u^{\infty}x{f_{1}\left(x\right)}{\rm{d}}x\\
{\sigma_t^2}&=L\left(\sigma^2+\left(u-\frac{\mu_t}{L}\right)^2\left(1-\frac{L}{N_{\rm{r}}}\right)\right),
\end{align}
\label{equ12}
\end{subequations}
where 
\begin{equation}
{\sigma^2}=\frac{N_{\rm{r}}}{L}\int_u^{\infty}x^2{f_{1}\left(x\right)}{\rm{d}}x,
\end{equation}  
and  $f_1(x)=e^{-x}$ denotes the PDF of $\tilde{\gamma}_{(i)}$. The constant $u$ satisfies $\int_{u}^{\infty}f_1(x){\rm{d}}x=\frac{L}{N_{\rm{r}}}$, and thus $u=\ln\frac{N_{\rm{r}}}{L}$. After substitutions and simplifications, the exact values of these two variables in Equ.(\ref{equ12}) reduce to
\begin{subequations}
\begin{align}
{\mu_t}&=L\left(1+\ln\frac{N_{\rm{r}}}{L}\right)\\
{\sigma_t^2}&=L\left(2-\frac{L}{N_{\rm{r}}}\right).
\end{align}
\label{equ14}
\end{subequations}

Since $\{\xi_h\}_{h=1,2,\cdots,N_{\rm{t}}}$ are i.i.d random variables, the asymptotic PDF of $\tilde{C}_{\rm{s}}$ termed as $p_{\rm{M}}(x)$ can be obtained by the characteristic function. Let $\Phi(j\omega)$ denote the characteristic function of $\tilde{C}_{\rm{s}}$, the asymptotic approximation of $\Phi(j\omega)$ reads
\begin{equation}
{\tilde{\Phi}(j\omega)}={\tilde{\Phi}_{\xi}^L(j\omega)},
\end{equation}  
where ${\tilde{\Phi}_{\xi}(j\omega)}$ represents the characteristic function of the asymptotic approximation for $\xi_h$. By Equ.(\ref{equ11}), ${\tilde{\Phi}_{\xi}(j\omega)}$ is written as
\begin{equation}
{\tilde{\Phi}_{\xi}(j\omega)}=\int_0^\infty{e^{j\omega\log_2\left(1+{\overline{\rho}}x\right)}\frac{1}{\sqrt{2\pi\sigma_t^2}}e^{-\frac{\left(x-\mu_t\right)^2}{2\sigma_t^2}}{\rm{d}}x}.
\label{equ16}
\end{equation} 
By substituting $t=\frac{x-\mu_t}{\sigma_t}$, $a=\frac{1+{\overline{\rho}}\mu_t}{{\overline{\rho}}\sigma_t}$ and $\zeta=\frac{j\omega}{\ln2}$ into Equ.(\ref{equ16}), ${\tilde{\Phi}_{\xi}(j\omega)}$ is reformulated as
\begin{equation}
\begin{aligned}
{\tilde{\Phi}_{\xi}(j\omega)}&=\frac{\left({\overline{\rho}}\sigma_t\right)^{\zeta}}{\sqrt{2\pi}}\int_{-\frac{\mu_t}{\sigma_t}}^{\infty}\left(t+a\right)^{\zeta}e^{-\frac{t^2}{2}}{\rm{d}}t\\&=\frac{\left({\overline{\rho}}\sigma_t\right)^{\zeta}}{\sqrt{2\pi}}F_{\zeta}.
\end{aligned}
\end{equation}
Thus the characteristic function of the asymptotic upper bound is ${\tilde{\Phi}(j\omega)}=\frac{\left({\overline{\rho}}\sigma_t\right)^{L\zeta}F_{\zeta}^L}{{2\pi}^{L/2}}$. It is necessary to perform an Fourier transform on ${\tilde{\Phi}(j\omega)}$ to acquiring the PDF of this asymptotic bound. Therefore, the PDF $p_{\rm{M}}(x)$ reads
\begin{equation}
p_{\rm{M}}(x)=\frac{1}{2\pi}\int_{-\infty}^{+\infty}\frac{\left({\overline{\rho}}\sigma_t\right)^{L\zeta}F_{\zeta}^L}{{2\pi}^{L/2}}e^{-j\omega{x}}{\rm{d}}\omega.
\label{equ18}
\end{equation}
Instead of direct integration, Equ.(\ref{equ18}) can be solved through Fast Fourier Transform (FFT) of sampling $\omega$. It is crystal clear that the sampling rate must be high enough to avoid aliasing.  

\section{Simulation Results}
\label{sec5}
In this part, simulation results are given for the former derivations. Sampling rate through $\omega$ for FFT is fixed to be 100 \rm{Hz} to prevent the aliasing phenomenon. The exact upper bounds in the following figures are all obtained through Monte-Carlo simulation consisting of a large number of experiments since the analytical forms of the upper bound are unknown. Times for experiments are set to be $5\times10^4$ to approach the exact upper bounds. Exact channel capacity for massive AS-MIMO system is obtained by exhaustive search when ${L}{\leq}{N_{\rm{t}}}$ due to its affordable hardware complexity. When $L$ is large, exhaustive search is prohibitive in complexity. Thus the capacity derived from greedy search (GS) \cite{b12}, which achieves near-optimal performance, can serve as alternative.

\begin{figure}[!t] 
\setlength{\abovecaptionskip}{-10pt} 
\centering 
\includegraphics[width=0.5\textwidth]{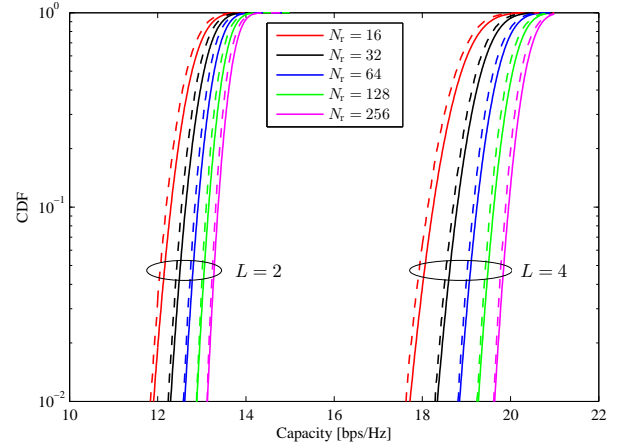} 
\caption{CDF of the asymptotic approximated BF upper bound and the exact upper bound, ${\overline{\rho}}=8{\text{dB}}$ and $N_{\rm{t}}=8$. The solid and dashed lines indicate the asymptotic approximated and exact distribution, respectively.} 
\label{figure4}
\end{figure}

\begin{figure}[!t]
\setlength{\abovecaptionskip}{-10pt} 
\centering 
\includegraphics[width=0.5\textwidth]{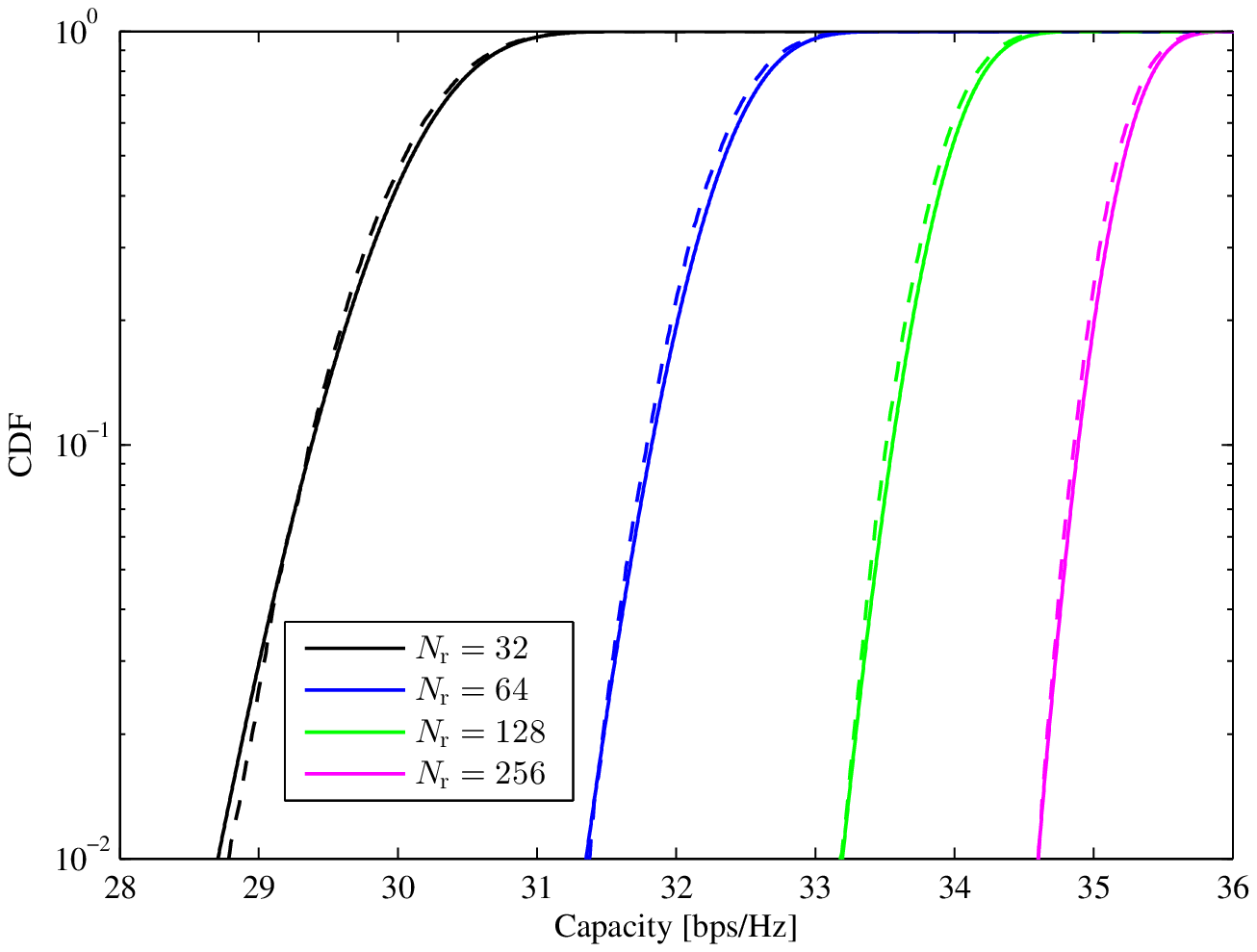} 
\caption{CDF of the asymptotic approximated MRC upper bound and the exact upper bound, ${\overline{\rho}}=8{\text{dB}}$, $N_{\rm{t}}=4$ and $L=20$. The solid and dashed lines indicate the asymptotic approximated and exact distribution, respectively.} 
\label{figure5}
\end{figure}

\begin{figure}[!t] 
\setlength{\abovecaptionskip}{-10pt} 
\centering 
\includegraphics[width=0.5\textwidth]{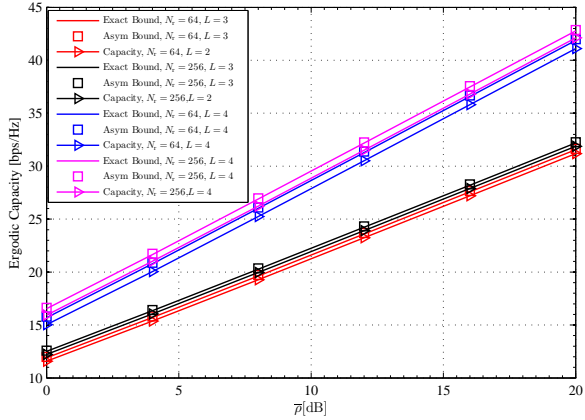} 
\caption{Ergodic capacity versus ${\overline{\rho}}$ when ${L}{\leq}{N_{\rm{t}}}$, $N_{\rm{t}}=8$. Asymptotic approximated bound, exact bound and channel capacity for AS-MIMO are denoted by Asym Bound, Exact Bound and Capacity, respectively.} 
\label{figure6}
\end{figure}

\begin{figure}[!t] 
\setlength{\abovecaptionskip}{-10pt} 
\centering 
\includegraphics[width=0.5\textwidth]{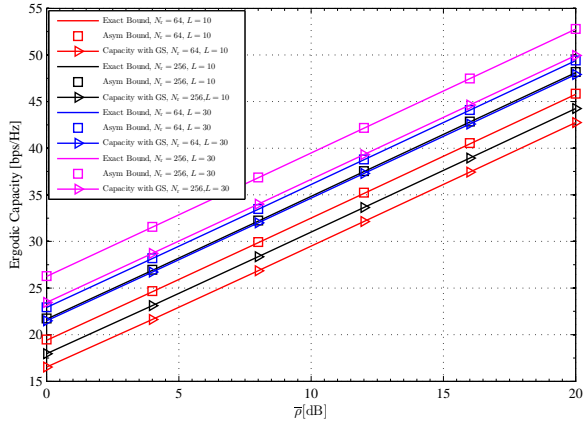} 
\caption{Ergodic capacity versus ${\overline{\rho}}$ when ${L}{>}{N_{\rm{t}}}$, $N_{\rm{t}}=4$. Asymptotic approximated bound, exact bound and channel capacity for AS-MIMO with greedy search are denoted by Asym Bound, Exact Bound and Capacity with GS, respectively.} 
\label{figure7}
\end{figure}

\begin{figure}[!t] 
\setlength{\abovecaptionskip}{-10pt} 
\centering 
\includegraphics[width=0.5\textwidth]{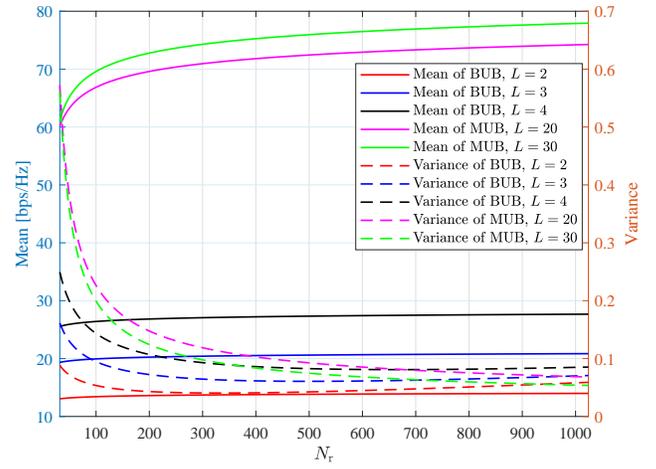} 
\caption{Mean and variance of asymptotic upper bound versus $N_{\rm{r}}$, $N_{\rm{t}}=8$ and ${\overline{\rho}}=8{\text{dB}}$.} 
\label{figure8}
\end{figure}

Fig.\ref{figure4} illustrates the cumulation distribution function (CDF) of the simulated upper capacity bound and the asymptotic approximated upper capacity bound for the massive AS-MIMO system when $L{\leq}N_{\rm{t}}$. Fix $\overline{\rho}=8{\rm{dB}}$ and $N_{\rm{t}}=8$, and the number of the selected antennas varies between 2 and 4 as $N_{\rm{r}}$ ranges from 16 to 256. The distribution of the exact upper bound is obtained by Monte-Carlo simulation as stated before and the asymptotic bound is calculated by Equ.(\ref{pB}). It is shown in \figurename \ref{figure4} that the CDF curves for the asymptotic approximated upper bound and the exact bound are almost coincident with the increase of $N_{\rm{r}}$. Furthermore, the asymptotic approximated bound has a fantastic approximation effect compared with the exact upper capacity bound even though $N_{\rm{r}}$ is at a moderate level, such as 32 and 64. Therefore, it makes sense to use the asymptotic theory to approximate the exact upper bound in massive MIMO systems.   

\figurename \ref{figure5} shows the CDF of the exact upper capacity bound by Monte-Carlo simulation and the asymptotic approximated upper capacity bound of MUB scenario when $L=20$, $N_{\rm{t}}=4$ and $\overline{\rho}=8{\rm{dB}}$. It is evident that the curves representing the asymptotic upper bound almost coincides with those of exact bound even though $N_{\rm{r}}$ is limited. It should be noted that asymptotic theory is an efficient and robust approximation tool for problems characterized by a large dimensionality, such as massive MIMO, which is intuitive in the light of the results exhibited in \figurename \ref{figure4} and \figurename \ref{figure5}. 

\figurename \ref{figure6} illustrates the ergodic capacity for the exact channel capacity and the BF upper bounds including both the asymptotic approximated one and the exact one. The ergodic value of the exact bound is obtained by Monte-Carlo simulation. It has been mentioned before that this upper bound is relatively tight when ${L}{\leq}{N_{\rm{r}}}$. From this figure, the ergodic values for asymptotic upper bound are nearlly equal to that of exact bound. Additionally, when $L$ is small, the BF upper bound is extremely tight according to the simulation results. When $L$ increases from 3 to 4, the bound becomes looser, which is consistent with the previous discussions in Section \ref{sec4.1}.

The ergodic capacity when ${L}{>}{N_{\rm{t}}}$ is plotted in \figurename \ref{figure7}. It is essentially unimplementable to obtain the accurate channel capacity when $L$ is large for the huge computation complexity. Nevertheless, greedy search can be used as benchmark in antenna selection instead for it can achieve near-optimal performance which reaches above 90\% of the optimal value according to the work in \cite{b9,b10,b11}. Actually, the capacity of greedy search is lower than the optimal value, but it is clear from the figure that the curves for the bound and greedy search is close, which means that the MUB is also relatively close with the exact channel capacity of AS-MIMO by ES. Moreover, the upper bound becomes tighter when $L$ gets larger, which verifies the demonstration in Section \ref{sec4.2} that the defined MRC bound is relatively tight when $L$ is large.

The mean and variance of the asymptotic approximated bound are illustrated in \figurename \ref{figure8}. It is shown that the variance converges to a tiny value gradually as $N_{\rm{r}}$ increases, which indicates that the bound will become more concentrated. The mean value gradually stabilizes or increases slowly as  $N_{\rm{r}}$ increases. These can be treated as the results of channel hardening effect \cite{b15,b31}. Such effects will be much more highlighted if the performance criterion is replaced with the exact channel capacity.

\section{Conclusion}
\label{sec6}
This paper studies the upper capacity bound for receive antenna selection in massive MIMO system. Asymptotic approximation for the bound is derived under the assumption that the number of receive antennas is boundless, which can be also applied when the total is at a moderate level. Simulation results show that the derived asymptotic bounds can achieve excellent performance. Furthermore, the experiments and comparison results show that the proposed upper bound is relatively tight in both MUB and BUB cases, which means upper bound can serve as a evaluation criteria for antenna selection in massive MIMO systems.

\vspace{12pt}
\end{document}